\documentclass{aa} 

\usepackage[varg]{txfonts}
\graphicspath{ {fig/} }
\usepackage{natbib}
\usepackage{graphicx}

\usepackage{hyperref}
\hypersetup{
        colorlinks=true,
    citecolor=blue,
    filecolor=blue,
    linkcolor=blue,
    urlcolor=blue,
    bookmarksopen=true
    }

\usepackage{pifont}
\newcommand{\cmark}{\ding{51}}
\newcommand{\xmark}{\ding{55}}

\begin{document}

    \title{Constraining the properties of the potential embedded planets in the disk around HD\,100546}
    \author{Max~Ackermann~Pyerin\inst{1} \and Timmy~N.~Delage\inst{1} \and Nicol\'{a}s~T.~Kurtovic\inst{1} \and Mat\'{i}as~G\'{a}rate\inst{1} \and Thomas Henning\inst{1} \and Paola~Pinilla\inst{1,2}} 
    \institute{Max-Planck-Institute for Astronomy, Königstuhl 17, 69117 Heidelberg, Germany
    \and Mullard Space Science Laboratory, University College London, Holmbury St Mary, Dorking, Surrey RH5 6NT, UK \\
    \newline Bachelor Project at the University of Heidelberg. Lead Author \email{MaxTec@gmx.de}, Group Leader \email{pinilla@mpia-hd.mpg.de}}
    \date{Received \_; accepted \_}
    
    \abstract    
        {
        The protoplanetary disk around the star HD\,100546 displays prominent substructures in the form of two concentric rings. Recent observations with the Atacama Large Millimeter/sub-millimeter Array (ALMA) have revealed these features with high angular resolution and have resolved the faint outer ring well. This allows us to study the nature of the system further.
        }
        {
        Our aim is to constrain some of the properties of potential planets embedded in the disk, assuming that they induce the observed rings and gaps.
        }
        {
        We present the self-calibrated $0.9\,$mm ALMA observations of the dust continuum emission from the circumstellar disk around HD\,100546. These observations reveal substructures in the disk that are consistent with two rings, the outer ring being much fainter than the inner one. 
        We reproduced this appearance closely with a numerical model that assumes two embedded planets. We varied planet and disk parameters in the framework of the planet-disk interaction code \href{http://fargo.in2p3.fr/}{\texttt{FARGO3D}} and used the outputs for the gas and dust distribution to generate synthetic observations with the code \href{https://www.ita.uni-heidelberg.de/~dullemond/software/radmc-3d/index.php}{\texttt{RADMC-3D}}.
        }
        {
        From this comparison, we find that an inner planet located at $r_1 = 13\,$au with a mass $M_1 = 8 M_{\rm{Jup}}$ and an outer planet located at $r_2 = 143\,$au with a mass $M_2 = 3 M_{\rm{Jup}}$ leads to the best agreement between synthetic and ALMA observations (deviation less than $3\sigma$ for the normalized radial profiles). To match the very low brightness of the outer structure relative to the inner ring, the initial disk gas surface density profile needs to follow an exponentially tapered power law (self-similar solution), rather than a simple power-law profile.
        }
        {}

    
    \maketitle

\section{Introduction}

    The formation of planets in protoplanetary disks is one of the main subjects of study in modern astrophysics. A very intriguing part of this topic is how newborn planets interact and actively shape the disk they are embedded in. This idea of planet-induced substructures is currently driven by recent observations with advanced telescopes, which have revealed that circumstellar disks can exhibit various prominent substructures, such as rings, gaps, asymmetries, or spirals (e.g., \citet{alma_2015, long2018, Andrews_2018, andrews2020, cieza2021}). Some of these substructures are thought to be indeed created by planet-disk interactions, rendering the observed features indicators of planet formation (e.g., \citet{pi1,pi2,zu3}), and may be used to constrain the properties of an embedded planet candidate, such as the mass and the orbital separation (e.g., \citet{ri}).
    
    The subject of this study is the circumstellar disk surrounding the Herbig star HD\,100546. This target is found at a distance of $108.1\pm 0.5\,$pc (\citet{gaia_edr3_2021}), with a right ascension of $11^h\, 33^m\, 25.3^s$ and a declination of $-70^\circ\, 11'\, 41.2''$ (\citet{gaia_edr3_2021}). Observations at multiple different wavelengths detected this disk, which spans $390\pm20\,$au outward from the star in CO emission and $230\pm20\,$au in continuum emission (\citet{w14}). The following substructures were revealed: a ring of emission with a radius of $27\,$au (Hubble Telescope, \citet{a7}), and a second very faint ring at $200\,$au (Atacama Large Millimeter/sub-millimeter Array - ALMA, \citet{w14}). 
    
    While several other mechanisms have been proposed to produce ring-like substructures in protoplanetary disks, such as the dead-zone outer edge mechanism (\citet{Pinilla_2016}), particle growth by condensation near ice lines (\citet{Zhang_2016,st17}), and magnetic disk winds (\citet{su193d}), in this work we assume the planet-disk interaction to be the main driver for the observed substructures of HD\,100546. In this mechanism, an embedded planet creates a pressure trap outward of its orbit that triggers an accumulation of dust by preventing the material from moving inward. This has been motivated in previous works, where simulations were executed to prove the similarity between structures created by planets and the observed structures of HD\,100546 (\citet{w14,montesinos2015,q15,p15,fedele2021}). For example, \citet{p15} estimated that a model with two high-mass planets can qualitatively reproduce the two rings observed in this disk. The inner planet is located at $10\,$au and the outer planet at $70\,$au. Furthermore, planet candidates were previously identified in this system by direct imaging (\citet{q13,q15}), but could not be confirmed because of confusion with disk features \citep{currie2014, currie2015}.
    
    In this study, our aim is to further constrain the properties of the potential embedded planets in the HD\,100546 circumstellar disk. Particularly, we wish to determine whether one or two planets are required to explain the observed substructures and to obtain their masses and distances from the central star. To do this, we calibrated and analyzed recent ALMA observations of the system to identify the main features and focused on resolving the outer ring that was suggested in \citet{w14,p15}. On the numerical side, we carried out hydrodynamic multifluid simulations (gas and dust) with \texttt{FARGO3D} (\citet{b16, m0}), and then generated synthetic observations from the outputs using \texttt{RADMC3D} (\citet{d12}). In this way, we tested different planet setup scenarios until we reached the best possible agreement between the synthetic and real observations by comparing their respective images and radial profiles.
    
    The layout of this paper is as follows. In Sect.~\ref{obs} we present the self-calibrated $0.9\,$mm ALMA observations of HD\,100546. In Sect.~\ref{sim} we describe the method of our hydrodynamic models and planet setups and also describe how synthetic observations can be obtained from them. In Sect.~\ref{psac} we present the synthetic observations that we compare with the ALMA observations. In Sect.~\ref{discussions} we discuss what can be inferred about the HD\,100546 system from these comparisons. Finally, Sect.~\ref{conclusions} summarizes our main findings.

\section{ALMA Observations} \label{obs}

    \subsection{Calibration}
        The data sets studied in this work include ALMA observations of HD\,100546 at $0.9$\,mm wavelength (ALMA Band 7) under ALMA projects 2016.1.00497.S. (PI: A. Pohl) and 2015.1.00806.S (PI: J. Pineda), to which we refer as short-baseline dataset (SB) and long-baseline dataset (LB), respectively. For the SB dataset, the project aimed to measure linear polarization in the emission, and therefore the correlator was configured to observe dust continuum in its four spectral windows centered at $336.495\,$GHz, $338.432\,$GHz, $348.494\,$GHz, and $350.494\,$GHz, each with a bandwidth of $2\,$GHz. The polarization setup of this project was tuned to measure the polarizations (XX, XY, YX, and YY), but we only used the total intensity information. For the LB dataset, the correlator was configured to observe two spectral windows with dust continuum at $331.988\,$GHz and $343.488\,$GHz, and the two remaining windows were aimed at the molecular lines $^{12}$CO ($J:3-2$) and $^{13}$CO ($J:3-2$). The bandwidth of the dust continuum spectral windows in the LB dataset was also $2\,$GHz and $0.469\,$GHz for the remaining windows. The details of all the data are summarized in Table~\ref{tab:obs_log}.
        
        After ALMA standard pipeline calibration, we used \href{https://casa.nrao.edu/index.shtml}{\texttt{CASA 5.6.2}} to handle and self-calibrate the datasets. In order to extract the dust continuum emission from the spectral windows with molecular line emission, we flagged the channels located at $\pm 25\,$km\,s$^{-1}$ from each targeted spectral line. Similarly to the calibration of the DSHARP sources \citep[see ][]{andrews2018}, the remaining channels from all spectral windows were averaged into 125\,MHz channels.
        
        The self-calibration of the SB data was made in two stages: First, we self-calibrated the shortest baseline observations, taken in April 2017, with solution intervals of $360\,$s, $150\,$s, $60\,$s, and $24\,$s for the phase calibration and $150\,$s for the amplitude calibration. This calibrated dataset was joined with the remaining observations from SB taken on October 2016 and was again self-calibrated by repeating the same solution intervals. The LB data were self-calibrated independently from the SB data, with solution intervals of $360\,$s and $150\,$s for the phase calibration and $360\,$s for the amplitude.
        
        Before combining the datasets, we corrected their phase centers such that the center of all observations was the center of the LB dataset. After self-calibration, we reduced the data volume by averaging the continuum emission into one channel per spectral window and $30$\,s of time binning. Finally, we applied the JvM correction to the images generated from the self-calibrated datasets. This correction accounts for the volume ratio $\epsilon_v$ between the point spread function of the images and the restored Gaussian of the {\small{\texttt{CLEAN}}} beam, as described in \citet{jorsater1995} and also used in \citet{andrews2021}. We find $\epsilon_v=0.69$ for the SB and LB images and $\epsilon_v=0.88$ for the images with SB and LB combined.

    \subsection{Observational results: Outer ring}

\begin{figure*}[t]
                \centering
                \includegraphics[width=0.96\linewidth]{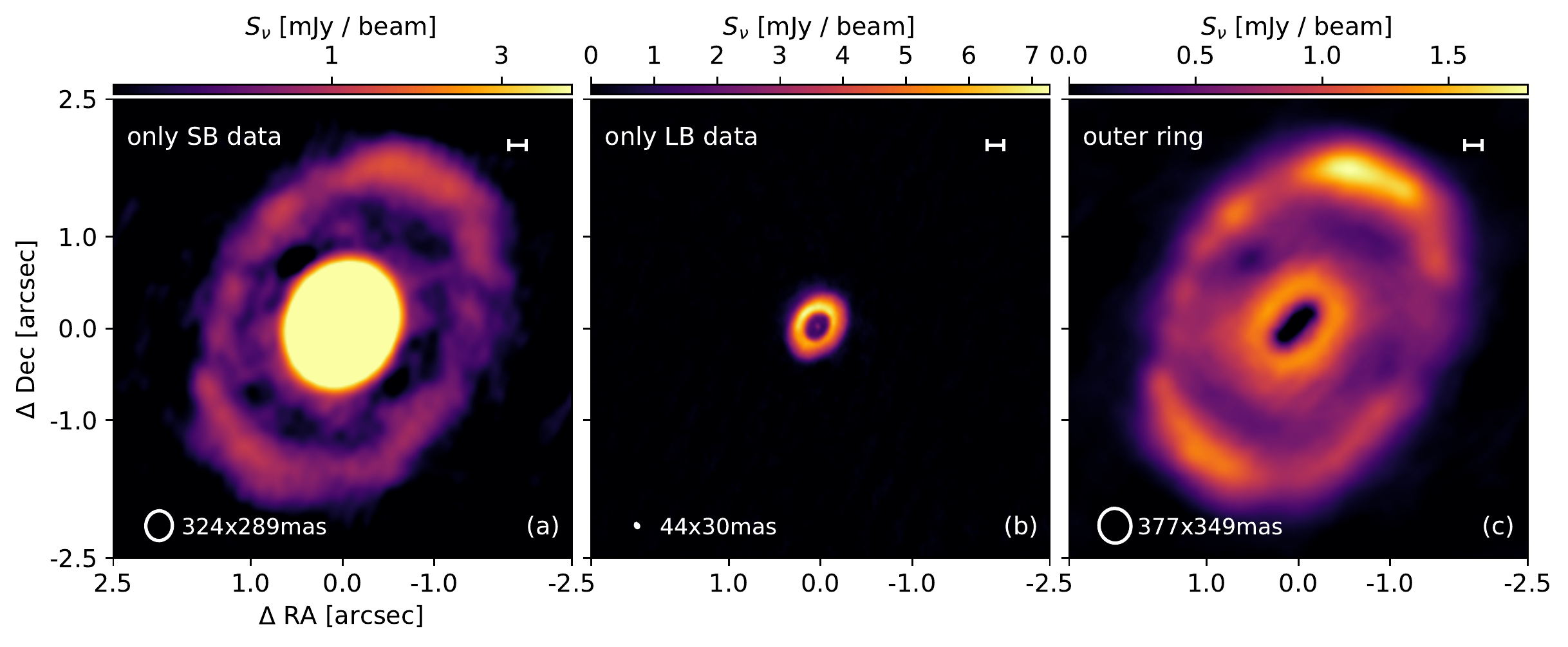}
                \caption{HD\,100546 protoplanetary disk dust continuum emission at $\lambda=0.9\,$mm. Panels a) and b) are the images generated using only the SB and LB datasets, respectively, with SB being on asinh scale and saturated to show the faint outer ring. Panel c) shows the outer ring structure when the SB and LB datasets are combined after subtraction of the visibilities from the inner disk. The scale bar represents 20\,au at the distance of the source, the lower left ellipse represents the beam size of each image, and the numbers to the right of the ellipse indicate the beam size in milliarcseconds. Panels a) and b) were generated with a robust parameter of 0.5, and panel c) was generated with a robust parameter of 1.8.}
                \label{im}
            \end{figure*}

        The inclusion of the different baselines allowed us to cover and resolve all the relevant scales of the disk. As shown in Fig.~\ref{im}, we detect the bright compact disk around HD100546 \citep[previously published in ][]{perez_2020}, as well as an extended faint ring that is only detectable when the SB information is included in the image reconstruction. When the disk is deprojected using the geometry from \citet{casassus_2019}, with an inclination of $45^\circ$ and a position angle (PA) of $150^\circ$, we find that the outer disk peaks at $\approx1.85''\pm0.1''\,$ from the disk center, which is $200\pm11\,$au at the distance of the source. Alternatively, we also imaged the outer structure after subtracting the contribution to the visibilities from the inner disk emission, in an attempt to reduce the high brightness dynamic range between the rings. This image is shown in panel c of Fig.~\ref{im}.

        To measure the flux from the outer ring, we calculated the flux density from the image generated using only the SB dataset with an elliptical mask with the inclination and position angle of the source, and we integrated the emission between $1.1''-2.5''$, which effectively masked the inner disk emission. We detect the ring to be $4.5\pm0.2\,$mJy, which does not include the $10\%$ ALMA flux uncertainty.
        
        We also estimated the optical depth $\tau$ of the emission to estimate the dust mass content in this region. We assumed that $\tau=-\ln(1 - T_{\text{bright}}/T_{\text{phys}})$, where $T_{\text{bright}}$ is the brightness temperature of a blackbody, and $T_{\text{phys}}$ is the physical temperature of the midplane, which we assumed to be $T_{\text{phys}}=20\,$K constant. The temperature profile of the outer ring peaks at $0.1\,$K, and this very low temperature translates into an equally low estimated optical depth of $\tau_{\text{peak}}=0.005$. This is highly uncertain, however, because the underlying estimate of $T_{\text{phys}}$ has a high level of uncertainty.
        
        The dust mass of the model was estimated by assuming optically thin emission and assuming that the flux ($F_\nu$) received at $0.9\,$mm is being emitted by dust with a constant temperature of $T=20\,$K \citep[as e.g.,][]{ansdell2016, pinilla2018}. We followed \citet{h83},
        \begin{equation}
            M_{\text{dust}} = \frac{d^2\,F_\nu}{\kappa_\nu\,B_\nu(T)} \text{,}
        \label{eq:mdust}
        \end{equation}
        \noindent where $d$ is the distance to the star, $\nu$ is the observed frequency, $B_\nu$ is the Planck function, and  $\kappa_\nu=2.3 \times (\nu/230\,\rm{GHz})^{0.4}\,\rm{cm}^{2}\rm{g}^{-1}$ is the frequency-dependent mass absorption coefficient \citep[as in][]{andrews2013}. 
        
        By using this $ F_\nu = 4.5\,$mJy for the faint outer ring in Eq.~\eqref{eq:mdust}, we obtained a dust mass of $0.7\pm0.07\,M_\oplus$, where we assumed a conservative uncertainty of $10\%$. The error of this estimate is, in addition to the uncertainty of the ALMA flux, mainly caused by the mass absorption coefficient, because this parameter depends on the dust grain size distribution and composition, which are unknown for protoplanetary disks. 
        
            \begin{figure}
                \centering
                \includegraphics[width=0.96\linewidth]{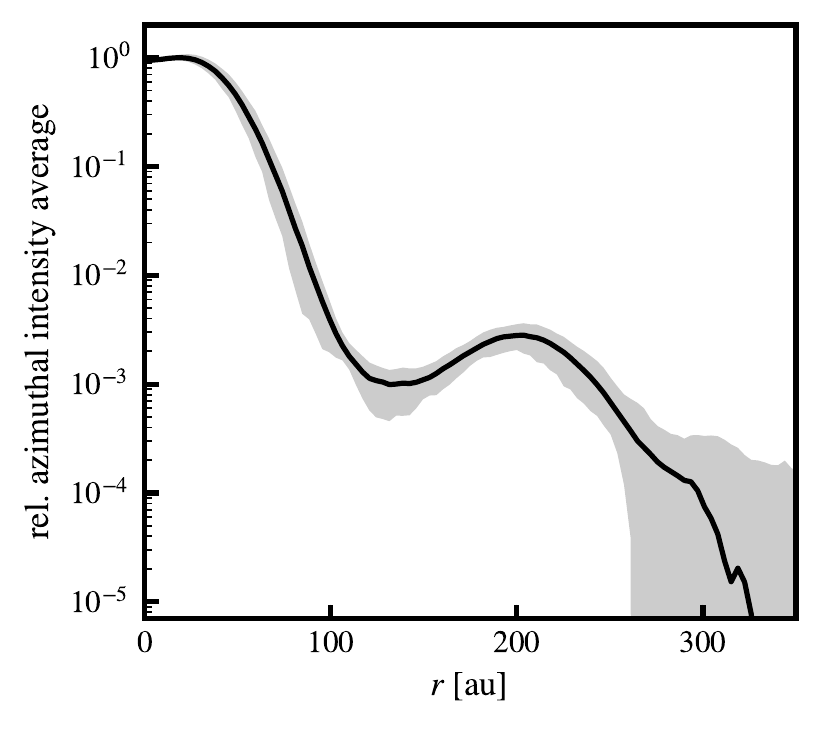}
                \caption{Radial profile corresponding to the ALMA observation shown in Fig.~\ref{im}. The gray interval marks the $1\sigma$ error obtained from statistical analysis of deviations in the image.}
                \label{rp}
            \end{figure}

\section{Numerical simulations} \label{sim}

    In this section we describe  our numerical approach and the underlying assumptions we used to model the HD\,100546 system. First,  we derive estimates for the locations and masses of the potential embedded planets in Sect.~\ref{planet_parameters}. Second, we describe our hydrodynamic simulations in Sect.~\ref{fargo_sim}, with the two initial conditions we explored and our modifications of the code to manually taper the mass of the planets over time (Sects. ~\ref{init} and \ref{timing_setup}, respectively). Third, we describe how synthetic observations can be obtained from the multifluid FARGO3D simulations in Sect.~\ref{radmc_synth_obs}. Finally, we outline our exploration of the parameter space in Sect.~\ref{parameterSEARCH}.
    
        \begin{table}
            \centering
                \caption{List of parameters used in the various simulations.}
                \begin{tabular}{lll} 
                    \hline
                    \hline
                    Parameter & Symbol/Units & Value \\
                    \hline
                    Star mass & $M_\star[M_{\odot}]$ & 2.13\\
                    Dust settling factor & $\chi$ & $3\%$\\
                    Inner disk boundary & $r_{\rm{min}}[\rm{au}]$ & 2.6 \\
                    Outer disk boundary & $r_{\rm{max}}[\rm{au}]$ & 500\\
                    Radial grid resolution & $n_r$ & 540 \\
                    Azimuthal grid resolution & $n_\phi$ & 1024\\
                    Critical radius & $R_c\,$[au] & 80\\
                    Intrinsic dust density & $\rho_{\rm{bulk}}[\rm{g}/\rm{cm}^3]$ & 1.5\\
                    Dust grain sizes & $a_{\rm{dust}}\,[\mu\rm{m}]$ & 0.1, 4.6, \\
                    &&220, $10^4$\\
                    Dust to gas mass ratio & $\epsilon$ & 1\%\\
                    Gas disk mass & $M_{\rm{disk}}[M_{\rm{Jup}}]$ & 21 \\
                    Dust disk mass & $M_{\rm{dust}}[M_\oplus]$ & 66 \\
                    \hline
                \end{tabular}
            \label{parametertable}
        \end{table}
    
    \subsection{Planet parameters} \label{planet_parameters}
    \subsubsection{Planet locations} \label{planet_setups}
        
        We deduce the orbit of the planet candidates by analyzing the radial profile of the combined ALMA image. To do this, we need to first understand how a massive planet affects the disk emission.
        
        If a planet is massive enough, it can open a gap in the gas distribution along its orbit. This creates a dust trap at the outer edge of the gap, where inward-drifting dust grains accumulate. This accumulation can then be observed as a ring (e.g., \citet{p15}). This ring is located outward of the planetary orbit. Additionally, the gas surface density decreases toward the position of the planet, and hence it is expected that the planetary orbit should correspond to the local minimum in the radial profile of the observed flux.
        
        From this assumption and the radial profile of our combined ALMA observations (Fig.~\ref{rp}), we find two promising locations for embedded planets: (1) The first location is such that it can explain the prominent inner substructure peaking at $27\,$au. Because we were unable to resolve the corresponding local minimum, we reverted to the estimated planet position from previous studies, specifically \citet{p15,w14}, suggesting that the inner planet is located at $13\,$au. (2) Our combined ALMA image also reveals a ring with a peak at radius $200\,$au, which suggests that another planet is needed to explain the observed emission. We estimate the orbital separation of the outer planet to be at $143\,$au, corresponding to the local minimum in the observed intensity profile.
        
        It is important to note that the assumption that exactly one planet is responsible for each gap is a simplification. Multiple planets with orbits close to each other could open a shared gap (e.g., as examined in the PDS70 disk, \citet{Bae_2019}).
        Furthermore, the behavior of opening a gap along the orbit and one ring being located outwards per planet is motivated by simulations of this specific disk. However, under different circumstances, a more complex behavior such as multiple rings induced by one planet and the planet positions coinciding with a dust ring has been found (\citet{Dong_2017,Dong_2018}).
        
    \subsubsection{Planet mass estimates} \label{planet_mass_estimates}
    
        The planet mass has a strong effect on the width of the induced gap. The precise relation of gap width and planet mass was studied by (\citet{Dodson_Robinson_2011, Pinilla_2012, rosotti2016, Fung_2016, 2018facchini}), yielding the dependence 
        
        \begin{equation}
            M_K = 3M_\star \left( \frac{D}{Kr} \right)^3\text{,} \label{mass_K_range_formula}
        \end{equation}
        
        \noindent where $M_K$ is the planet mass estimate, which depends on the dimensionless parameter $K$, the stellar mass $M_\star$, the orbit of the planet $r$, and the gap width in the millimeter dust continuum $D$ (defined following \citet{l19} as the minimum to peak radial distance), obtained from observations. $K$ describes how many Hill radii $R_{\rm{Hill}}=r\sqrt[3]{M/3M_\star}$ of the planet correspond to the continuum gap width. \textit{K}\textit{} has been found to lie within the interval $7\leq K \leq 10$. This allows us to find mass estimates directly from the radial profile of the dust continuum emission (Fig.~\ref{rp}). We inferred the continuum gap width as the radial distance of the local flux peak of the substructure to the corresponding estimate of the planet position. This yields $D_1=14$ au for the inner and $D_2=79$ au for the outer gap, which corresponds to the mass estimates $8\,M_{\rm{Jup}}\leq M_{K,1} \leq 24\,M_{\rm{Jup}}$ for the inner and $1.1\,M_{\rm{Jup}} \leq M_{K,2} \leq 3.3\,M_{\rm{Jup}}$ for the outer planet.
        
        We also took the minimum planet mass $M_{\rm{min}}$ into account that is required to open a gap at all, following the studies of \citet{c6} and \citet{l19}. First, there is the model of \citet{c6}. This model is a balance between the pressure and viscosity torque from the disk with the gravitational torque from the planet. The authors estimated the minimum mass to be 
        
        \begin{equation}
            M^{\rm{\,Crida}}_{\rm{min}} = \left\{ qM_\star \left| \frac{3 \sqrt[3]{3} H }{4 r} q^{-1/3} + \frac{50 \alpha c_s H}{\sqrt{\frac{G M_\star}{r^3}} r^2} q^{-1} = 1 \right.\right\}\text{,}
            \label{CridaFormula}
        \end{equation}

        \noindent where $c_s=\sqrt{\frac{k_B T}{\mu m_H}}$ is the isothermal sound speed, $H = \frac{c_s}{\Omega_K}$ is the disk pressure scale height, $q$ is the planet mass in units of the stellar mass, and $\alpha$ is the disk viscosity from the $\alpha$-disk model (\citet{s73}). Then there is the model of \citet{l19}, which is based on hydrodynamical simulations that include dust dynamics. It finds a minimum mass of
        
        \begin{equation}
            M^{\rm{\,Lodato}}_{{\,min}} = 
                \begin{cases}
                    \, 0.3 M_\star \left(\frac{H}{r}\right)^3 & \text{for} \, \text{St} < 1 \\
                    \, 3 M_\star \left( \frac{H}{5.5 r \sqrt{\text{St}}} \right)^3 & \text{for} \, \text{St} \geq 1\text{,}
                \end{cases}
                \label{LodatoFormula}
        \end{equation}
        
        \noindent where St corresponds to the the Stokes number, defined by 
        
        \begin{equation}
            \text{St} = \frac{\pi \, a_{\rm{dust}} \, \rho_{\rm{bulk}} }{2\, \Sigma_{g}}\text{,}
        \end{equation}
        
        \noindent with $a_{\rm{dust}}$ being the dust grain size, $\rho_{\rm{bulk}}$ the intrinsic dust density, and $\Sigma_{g}$ the gas surface density.
        
        From these two criteria, we estimated the minimum planet mass by choosing the maximum of the two models,
        
        \begin{equation}
            M_{\rm{min}} = \rm{max}\left(M^{\rm{Crida}}_{\rm{min}}, M^{\rm {Lodato}}_{{\,min}}\right).
        \end{equation} 
        
        \noindent For $\alpha=10^{-4}$, we obtain $M_{\rm{min},1}=0.3\,M_{\rm{Jup}}$ for the inner and $M_{\rm{min},2}=1.7\,M_{\rm{Jup}}$ for the outer planet. For $\alpha=10^{-3}$, the minimum mass estimates are $M_{\rm{min},1}=0.7\,M_{\rm{Jup}}$ for the inner and $M_{\rm{min},2}=3\,M_{\rm{Jup}}$ for the outer planet. 
        
        We took the lowest possible mass for each of the planets for our simulations because planets have not directly been observed in this disk, and thus a lower-mass companion is more likely to exist. Therefore we determine the planet masses $M$ for the \texttt{FARGO3D} setups as
        
        \begin{equation}
            M = \rm{max}\left(\rm{min}_K\left(M_K\right), M_{{\,min}}\right)\text{,}
            \label{mass_master_model_eq}
        \end{equation}
        
        \noindent which returns an estimate of $M_1=8\,M_{\rm{Jup}}$ for the inner planet mass. For the outer companion, we obtain different estimates depending on $\alpha$: $\alpha=10^{-4}$ leads to $M_2=1.7\,M_{\rm{Jup}}$, and $\alpha=10^{-3}$ implies $M_2=3\,M_{\rm{Jup}}$.

    \subsection{Hydrodynamic simulations with \texttt{FARGO3D}} \label{fargo_sim}

        We modeled this disk with the numerical hydrodynamics solver \href{http://fargo.in2p3.fr/}{\texttt{FARGO3D}} (\citet{b16, m0,2019fargomulti,2019diffusionfargo}), which  is designed to simulate planet-disk interactions. This code evolves a numerical fluid over time that is described with a grid of densities, velocities, and energies.
        
        We used a multifluid approach and model the gaseous dusty disk around HD\,100546 as five fluids: one fluid represents the gas, and the four remaining fluids describe the dust, where each of them corresponds to a specific grain size $a_{\rm{dust}} \in \left\{0.1\,\mu\rm{m}, 4.6\,\mu\rm{m}, 0.22\,\rm{mm},\text{ and } 1\,\rm{cm} \right\}$.

        The \texttt{FARGO3D} simulations were performed in two dimensions on a polar grid $(r,\varphi)$, extending azimuthally from $0$ to $2\pi$ and radially from $r_{\rm{min}}$ to $r_{\rm{max}}$. The radial extent was determined in order to enclose the orbits of all planet candidates as well as the induced substructures without any potential boundary effects. They were calculated based on the planet Hill radius $R_H$ as  $r_{\rm{min}}=(r_1-3 R_{H,1})/3 =2.6\,$au and $r_{\rm{max}}=3(r_2+3 R_{H,2})=500\,$au. For the radial simulation boundaries, we used a power-law extrapolation as the boundary condition for the densities and standard Keplerian speed/antisymmetric boundaries for the velocities in the $r$/$\varphi$ direction ({\small\texttt{KEPLERIAN2DDENS}} and {\small\texttt{KEPLERIAN2DVAZIM}}/{\small\texttt{ANTISYMMETRIC}} in the \texttt{FARGO3D} setup, respectively). The grid resolution was chosen such that one grid cell corresponded to one-tenth of the pressure scale height H at the two planet locations $r_1 = 13\,$au and $r_2 = 143\,$au, which equates to an azimuthal resolution of $n_\phi = 1024$ cells and a radial resolution of $n_r = 540$ cells.
        
        To determine the duration of the simulations, we need to consider the age of the HD\,100546 system, which has been estimated to be $4.8\,\rm{Myr}$ (\citet{Wichittanakom_2020}). However, a simulation of such a duration is computationally unfeasible because we need to choose a high grid resolution to resolve the disk at the locations of the two planet candidates. We instead ran our simulations for $ 0.74\,$Myr (corresponding to $2.3\times 10^4$ orbits of the inner planet or $6.3\times 10^2$ orbits of the outer planet), assuming that, qualitatively, the disk reaches a quasi-steady state within this time. We discuss this assumption in Appendix \ref{conv_behav}.

    \subsubsection{Initial conditions} \label{init}
        
        To model the initial distribution of the gas surface density $\Sigma_{g}$ of the disk for the \texttt{FARGO3D} simulations, we used two approaches. The first approach was a simple power-law profile,
        
            \begin{equation}
                \Sigma_g (r,\varphi) = \Sigma_0 \left(\frac{r}{r_1}\right)^{-1}\text{,}
                \label{pow_eq}
            \end{equation}

        \noindent where $r_1=13$ au is the orbit of the innermost planet, which was used as a length scale by \texttt{FARGO3D}, and $M_{\rm{disk}}$ is the total gas disk mass enclosed between $r_{\rm{min}}$ and $r_{\rm{max}}$. The normalization constant $\Sigma_0$ is obtained such that $M_{\rm{disk}} = \int_{r_{\rm{min}}}^{r_{\rm{max}}} \; 2 \pi r \Sigma_{g}(r) \; dr$, where we assumed the total gas disk mass to be $M_{\rm{disk}} = M_{\rm{dust}}/\epsilon$. Here, $\epsilon = 10^{-2}$ corresponds to the dust-to-gas mass ratio, taken to be equal to the interstellar value, and $M_{\rm{dust}}=66\,M_\oplus$ is the total dust disk mass obtained from the total observed flux of our combined ALMA dust continuum emission of $435$ mJy (see Eq.~\eqref{eq:mdust}).
        
        The other initial condition used in this work is the self-similar solution from \citet{lynden-bell_pringle_1974},
        
            \begin{equation}
                \Sigma_g (r,\varphi) = \Sigma_0 \left(\frac{r}{R_c}\right)^{-1} \exp\left(-\frac{r}{R_c}\right)\text{.}
                \label{exp_eq}
            \end{equation}
        
        \noindent This profile is a power law combined with an exponential decay to reduce the amount of material in the outer disk. Here, $R_c = 80\,$au is the critical radius, and $\Sigma_0$ is chosen analogously to the above.
        
        The initial dust surface densities were set following $\Sigma_{d}(a_{\rm{dust}}) = f(a_{\rm{dust}}) \epsilon \Sigma_{g}$, such that the dust-to-gas ratio radial profile was radially constant and equal to $\epsilon$. The factor $f(a_{\rm{dust}})$ determines the mass distribution across the different dust sizes, and it was set such that the number density of the dust behaves as $n(a_{\rm{dust}}) \,da_{\rm{dust}} \propto a_{\rm{dust}}^{-3.5}\, da_{\rm{dust}}$, following an MRN distribution (\citet{mathis_1977}).

    \subsubsection{Mass-tapering model} \label{timing_setup}
        
        To mimic the fact that planets grow over time and avoid any numerical artifacts right at the beginning of the simulations by introducing the whole planet mass, we used the following mass-tapering approach. This formula is a variation of the mass-tapering algorithm that is included in \texttt{FARGO3D}. We changed the code because we needed to be able to set a different taper for each planet individually in order to study the effect of planet formation timescales. We also added a new offset parameter. Our model reads
        
            \begin{equation}
                \frac{m_i(t)}{M_i} = 
                    \begin{cases}
                        \,0 & \text{for } t<\Delta_i\\
                        \,\frac{1}{2} \left(   1 - \cos \left(  \pi\, \frac{t-\Delta_i}{\tau_i} \right)  \right) & \text{for } \Delta_i\leq t<\Delta_i+\tau_i \\
                        \,1 & \text{for } t \leq \Delta_i+\tau_i \text{,}
                    \end{cases}
            \label{masstaper}
            \end{equation}
        
        \noindent where  $m_i(t)$ is the effective mass of the planet $i$ at a given time $t,$ and $M_i$ is the final planetary mass. This model has two parameters, which we refer to as the "timing parameters": (1) $\Delta_i$, the time at which the planet is introduced into the simulation; and (2) $\tau_i$, the formation timescale after the planet is introduced. We fiducially set the delays $\Delta_1$ and $\Delta_2$ to zero and the tapers $\tau_1$ and $\tau_2$ to $16\,$Kyr. We chose $16\,$Kyr, corresponding to $1000$ orbits of the inner planet, as an arbitrary timescale that is small compared to the simulated duration, but still long enough to prevent numerical artifacts by introducing the full planet mass from the beginning of the simulation.
        
        We note that while our approach does increase the planet mass over time, the actual physical accretion process is ignored in this present work, that is, there is no mass transfer from the dust or gas fluids onto the planet. We also fixed the planets orbit and neglected its potential migration.

    \subsection{Synthetic observations with \texttt{RADMC-3D}}  \label{radmc_synth_obs}
    
        To directly compare the density distributions from \texttt{FARGO3D} with our self-calibrated ALMA image, we performed a radiative transfer simulation and rendered a synthetic observation, using the code \href{https://www.ita.uni-heidelberg.de/~dullemond/software/radmc-3d/index.php}{\texttt{RADMC-3D}} (\citet{d12}). This requires expanding the two-dimensional dust surface density distributions into three dimensions, which we did by using a Gaussian distribution along the vertical ($z$) axis, assuming hydrostatic equilibrium in this direction,
        
        \begin{equation}
            \rho_{d}(r,\varphi,z) = \Sigma_{d} (r,\varphi) \frac{1}{\sqrt{2\pi}H_{\rm{dust}}(r)}\exp\left(-\frac{z^2}{2 H_{\rm{dust}}(r)^2}\right) \text{,}
        \end{equation}

        \noindent with the disk height model of \citet{d4},
        
        \begin{equation}
            H_{\rm{dust}}(r) =
            \begin{cases}
                \,\chi\,H(r) & \text{for } a_{\rm{dust}} > 1\,\rm{\mu m} \\
                \,H(r) & \text{otherwise} \text{,}
            \end{cases}
            \label{diskheightmodel}
        \end{equation}
        
        \noindent where $H = \frac{c_s}{\Omega_K}$ is the disks pressure scale height, and $\chi=3\%$ is the scale height reduction factor encoding the effect of dust settling, which is more prominent for the larger grain sizes. It therefore results in a shorter dust scale height compared to the one for smaller grain sizes that roughly follow the gas.
        
        We assumed the disk to be composed of silicates (\citet{d95}) and used corresponding optical constants from the \href{http://www.astro.uni-jena.de/Laboratory/Database/databases.html}{Jena database}. To match our combined ALMA observations, the wavelength was fixed at $\lambda=0.9\,$mm, and all synthetic observations are convolved with a $0.3''$ Gaussian beam similar to the actual observations.

    \subsection{Parameter space exploration} \label{parameterSEARCH}
    
        \begin{table*}
            \centering
                \caption{Our model setups. The setup that agrees best with the observation is model VII.}
                \begin{tabular}{c|c|c|c|c|c|c|c} 
                    \hline
                    \hline
                    Model setup  &  I & II  & III & IV & V & VI & VII\\
                    \hline
                    Initial Density Model & pow (Eq.~\ref{pow_eq}) & exp (Eq.~\ref{exp_eq}) & pow (Eq.~\ref{pow_eq})& exp (Eq.~\ref{exp_eq})& exp (Eq.~\ref{exp_eq})& exp (Eq.~\ref{exp_eq})& exp (Eq.~\ref{exp_eq})\\
                    Disk Viscosity $\alpha$ &  $10^{-4}$ &  $10^{-4}$&  $10^{-4}$&  $10^{-4}$&  $10^{-4}$&  $10^{-4}$&  $10^{-3}$ \\
                    \hline
                    Inner Planet? & \cmark& \cmark& \cmark& \cmark& \cmark& \cmark& \cmark \\
                    Planet mass $M_1$ [$M_{\rm{Jup}}$] & $8$ & $8$ & $8$ & $8$ & $8$ & $8$ & $8$ \\
                    Delay Time $\Delta_1$ [Kyr] & $0$ & $0$ & $0$ & $0$& $0$ & $0$ & $0$\\
                    Mass Taper $\tau_1$ [Kyr] &  $16$ & $16$ & $16$ & $16$ & $16$ & $16$ & $16$ \\
                    \hline
                    Outer Planet? & \xmark & \xmark & \cmark& \cmark& \cmark& \cmark& \cmark\\
                    Planet mass $M_2$ [$M_{\rm{Jup}}$] & - & - & $1.7$ & $1.7$ &$1.7$ &$1.7$ & $3$ \\
                    Delay Time $\Delta_2$ [Kyr] & -& -& $0$ & $0$& $600$ & $0$ & $0$\\
                    Mass Taper $\tau_2$ [Kyr] &- &- & $16$ & $16$ & $16$ & $600\,$ & $16\,$ \\
                    \hline
                    \hline
                \end{tabular}
                
            \label{model_setup_table}
        \end{table*}
        
        In order to find a model that reproduces the ALMA observations, we ran multiple simulations with different parameters (see Table~\ref{model_setup_table}). We started by testing how models with a single planet compare to the observation, and ran a simulation for both the initial gas surface density profiles we introduced in Sect.~\ref{init} (model  I with the power law, and model II with the self-similar solution). Then, we executed simulations with two planets for both initial conditions (model III with the power law, and model IV with the self-similar solution). In all of these four simulations, we introduced the planet(s) right from the beginning of the run and used the fiducial taper of $\tau=16\,$Kyr as well as a fiducial viscosity of $\alpha=10^{-4}$.
        
        Then, we explored the effect of different planet formation timescales. Because the ALMA observation reveals that the inner ring is much brighter than the outer ring, the inner pressure trap must exist very early in order to trap as much material as possible. For this reason, we always introduced the inner planet immediately at the beginning of the simulations, that is, we always set the offset $\Delta_1 = 0$ and chose a fiducial mass taper of $\tau_1=16\,$Kyr, and we only varied the timing parameters for the second planet. In model V we set $\Delta_2 = 6\times 10^5\,$yr while keeping $\tau_2 =16\,$Kyr, while in model VI, we kept $\Delta_2 = 0$ and set $\tau_2 = 6\times 10^5\,$yr. This means that in models V and VI, the outer planet reaches its final mass on a timescale comparable to the duration of the entire simulation ($7.4\times 10^5\,$yr). In these simulations, the viscosity remained the same as for the previous ones for comparability ($\alpha=10^{-4}$), and the initial profile is the self-similar solution in both of them because we show below that such a profile is required to reach the best agreement between synthetic and real observations.
        
        Finally, we simulated a disk with a different viscosity $\alpha$. We chose a higher viscosity of $\alpha=10^{-3}$ in model VII. This simulation used the fiducial timing setup, that is, it introduced both planets from the start of the simulation with the fiducial taper ($\Delta_1=\Delta_2=0$, $\tau_1=\tau_2=16\,$Kyr) and used the the self-similar solution initial profile. It therefore directly compares to model IV. In this model, the mass estimate for the outer planet increases to $3\,M_{\rm{Jup}}$.
        
        The parameters that are common in all simulations are listed in table~\ref{parametertable}. An overview of the exact values of the parameters that differed between the seven different simulations is given in table~\ref{model_setup_table}.

\section{Results} \label{psac}

\begin{figure*}
            \centering
                \includegraphics[width=0.96\linewidth]{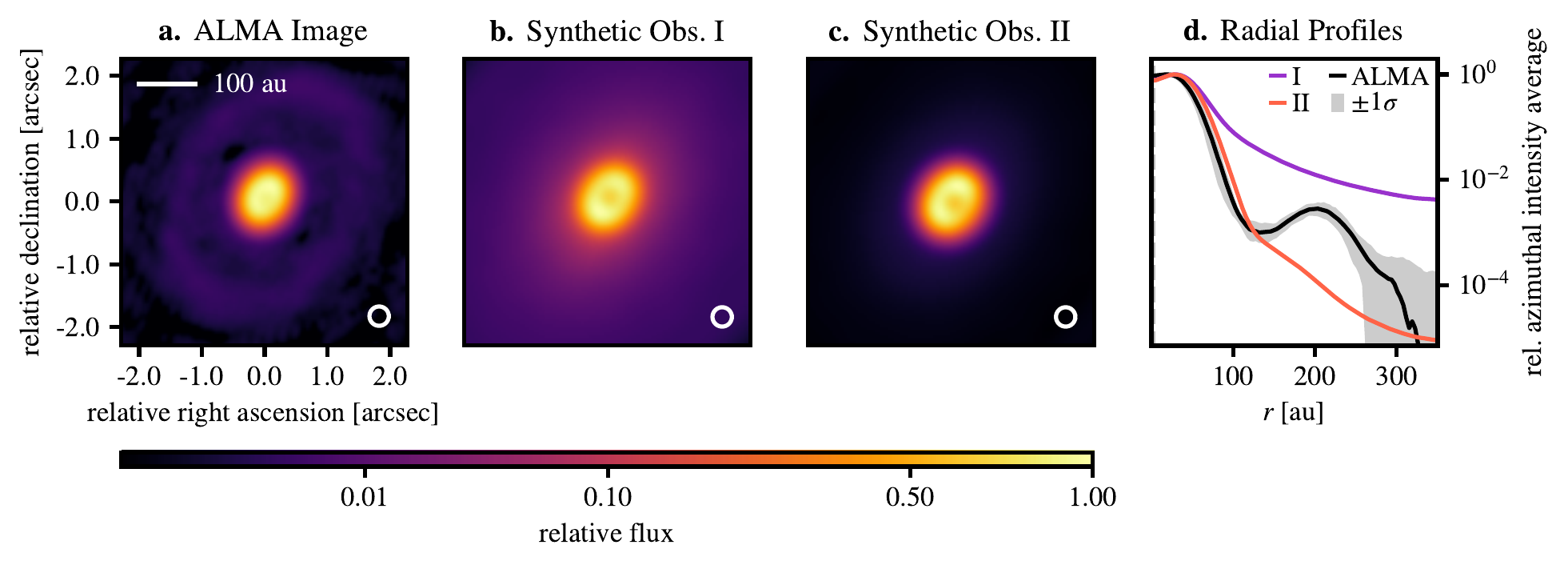}
                \caption{Results for the simulations with a single planet, positioned at $13\,$au with a mass of $8 M_{\rm{Jup}}$  (position marked in panel d as a dashed vertical line). The two simulation results correspond to different initial density models. Simulation I started with a power law (Eq.~\eqref{pow_eq}) for the radial fall-off of $\Sigma_{g/d}$ and simulation II with a self-similar solution profile (Eq.~\eqref{exp_eq}). We present them as synthetic observations convolved with a $0.3''$ beam (panels b and c) in direct comparison to the modeled ALMA image (panel a).}
                \label{im_1p}
            \end{figure*}

\begin{figure*}
                \includegraphics[width=0.96\linewidth]{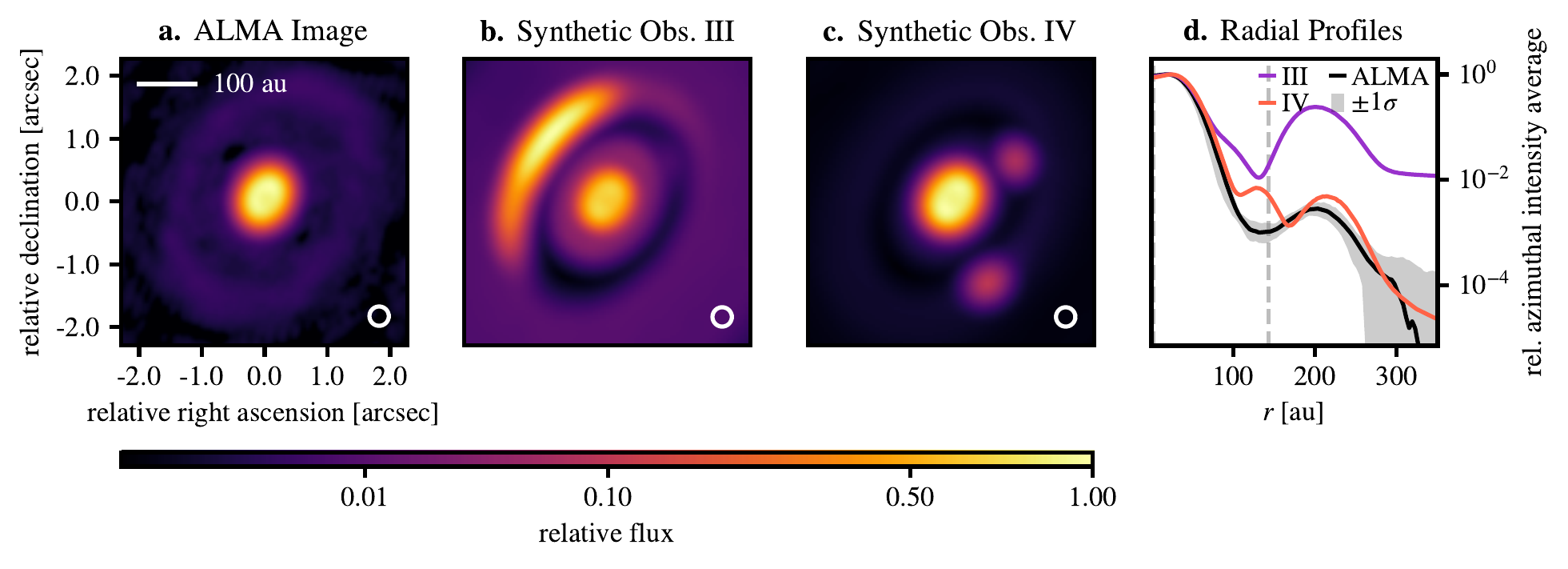}
                \caption{Outcome of the simulations including two planets. We positioned the inner planet at $13\,$au  with a mass of $8 M_{\rm{Jup}}$ and the outer planet at $   143\,$au with $1.7 M_{\rm{Jup}}$ (positions marked in panel d as dashed vertical lines). The planets form at the same time and equally fast, but the initial density profile is assumed differently in the two compared setups: the initial density profile for simulation III was a power-law, while simulation IV started as a self-similar solution profile. The inconsistent bump, shown as the red curve centered on the position of the outer planets, is an artifact of the lacking accretion process in the simulation, as can be concluded based on the results of \citet{b20}, who have shown the effeact of accretion on the gap formation process. They found that when the accretion process is included in the simulation, the gap is cleared out completely for a disk setup like ours. The accumulation at the position of the planet would therefore not be observed in reality.}
                \label{im_stf}
            \end{figure*}

\begin{figure*}                     
            \includegraphics[width=0.96\linewidth]{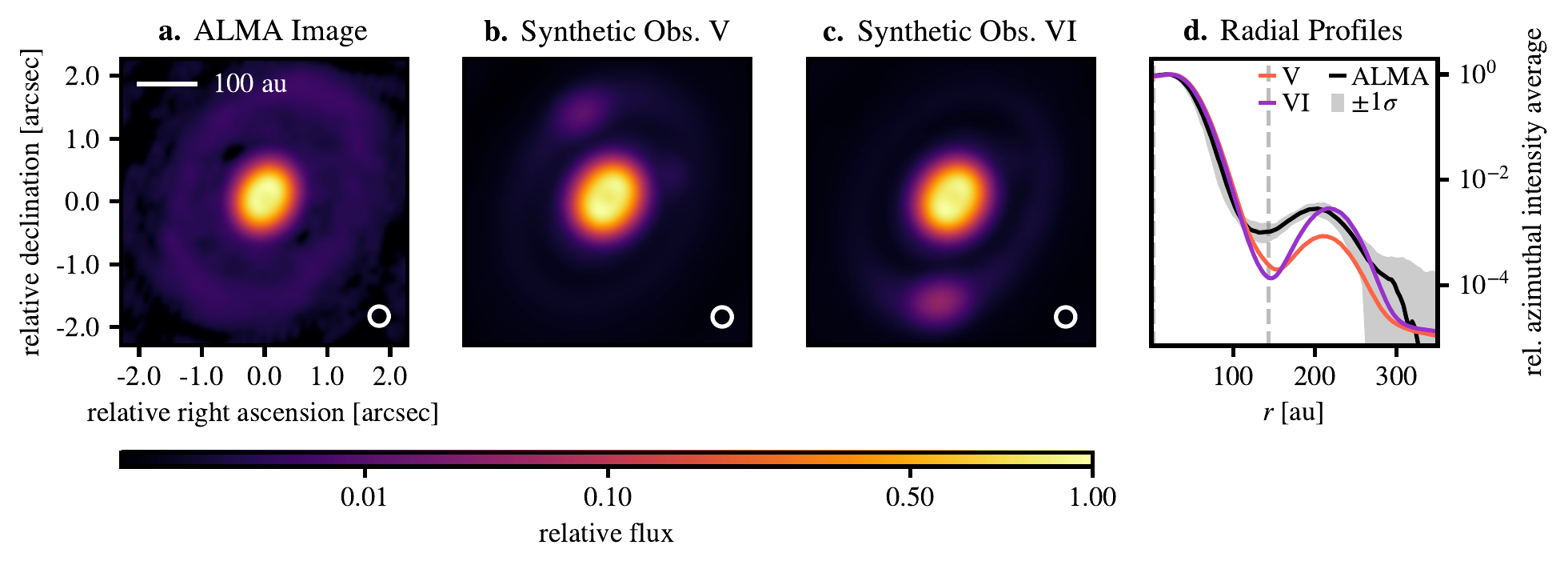}
                \caption{Comparison of the effect of different planet formation timescales. The planet setup is the same for both: Two planets, positioned at $  13\,$au and $   143\,$au, with a mass of $8 M_{\rm{Jup}}$ and $1.7 M_{\rm{Jup}}$. In simulation V, the outer planet started to form much later than the inner planet ($\Delta_1 = 0\,$, $\Delta_2 = 600\,$Kyr, $\tau_1 = \tau_2 = 16\,$Kyr). For simulation VI, the planets started to form simultaneously, but the outer planet gained mass significantly slower ($\Delta_1 = \Delta_2 = 0$, $\tau_1= 16\,$Kyr, $\tau_2 = 600\,$Kyr).}
                \label{im_time}
            \end{figure*}

\begin{figure*}
                \includegraphics[width=0.96\linewidth]{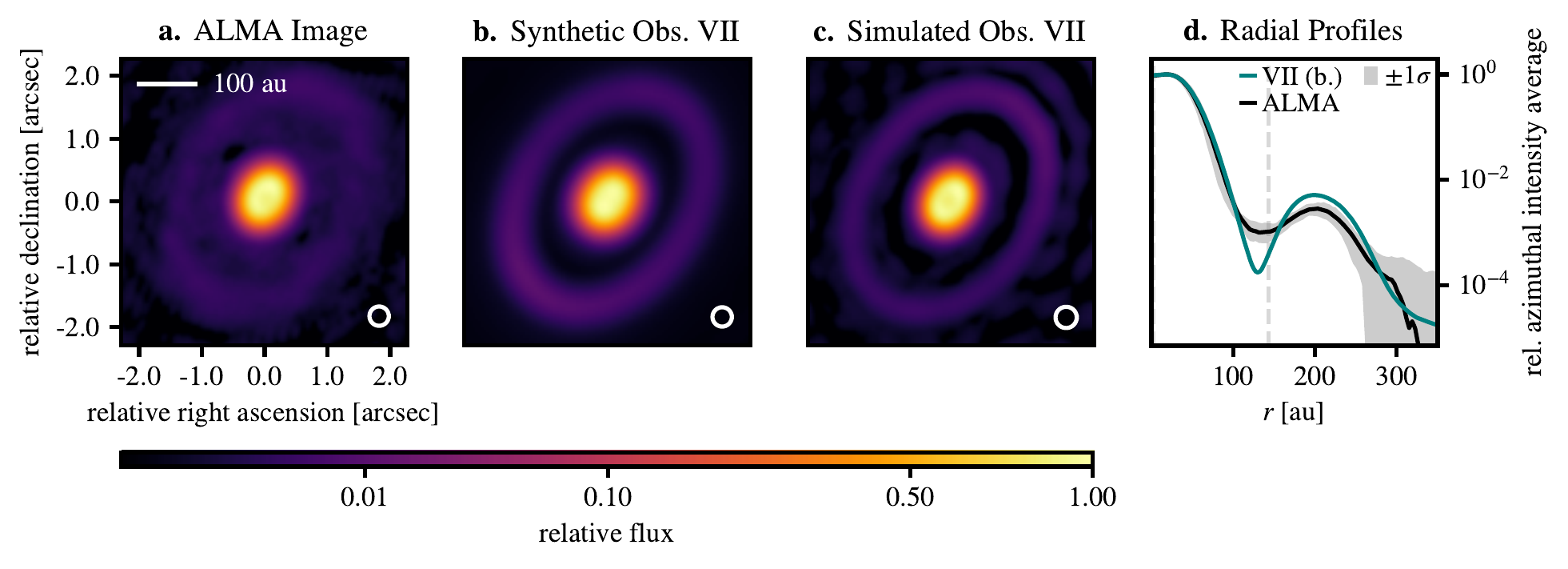}
                \caption{Result for a disk viscosity of $\alpha=10^{-3}$ (model setup VII). The mass estimates in this simulation are $8 M_{\rm{Jup}}$ for the inner and $3 M_{\rm{Jup}}$ for the outer planet. Because this setup shows the best agreement with the modeled ALMA image of all our results, we further simulated an observation with ALMA based on the unconvolved \texttt{RADMC-3D} output to proove the observability of the simulated substructures.  This simulated observation is shown in panel c of this figure.}
                \label{im_a3}
            \end{figure*}

        The main results we obtained for the different planet setups and disk parameters are shown in Figs.~\ref{im_1p}-\ref{im_a3}. For each setup, we present the corresponding synthetic observation and the combined ALMA image, both normalized to their respective maximum, and using the same color map. In addition, we display the azimuthally averaged normalized radial profiles as a metric to compare the two images quantitatively.

        Figure~\ref{im_1p} shows models I and II, which solely include the inner planet, with different initial density profiles (Eqs. \ref{pow_eq} and \ref{exp_eq}, respectively). Both of these models induce a ring in the synthetic observation that matches the inner ring of the real observation in the image and the radial profile. However, these models do not create an outer ring because there is no second planet that might induce an outer trap. Comparison of the radial profiles reveals how strongly the synthetic observations deviate from the observation. Model I reaches its largest deviation at $110\,$au with $133\sigma$, while it is $4\sigma$ at $200\,$au  for model II.
        
        In Fig.~\ref{im_stf} we show simulations including two planets (models III and IV) with different initial density models. In contrast to the one-planet models (I and II), both of these models create two rings. The location of these rings matches the ALMA observation, and the radial profiles show a similar shape. We therefore continued our parameter search with two planets.
        
        In model III the outer structures are relatively bright compared to the inner ring, which is strongly overestimated: in the model, the relative peak intensity of the outer ring is $86$ times higher than in the ALMA observation. This issue does not arise with model IV, which reproduces the relative peak intensity of the two substructures properly. Model IV is within a $3\sigma$ range around the observation in the normalized radial profile. We therefore continued our parameter search using two planets and the self-similar solution as initial condition.
        
        However, model IV presents an issue regarding the symmetry of the outer features: the ring is mainly concentrated in one point and is much fainter and thinner along the remaining structure in the azimuthal direction. This differs from the observed outer ring, which is evenly bright and symmetric. To resolve this issue, we conducted further tests to quantify the potential effect of different model parameters, specifically, the timing parameters and the disk viscosity $\alpha$.
        
        Figure~\ref{im_time} shows the results with different planet-mass tapering. The models are identical to model IV, except for one of the parameters of the mass-tapering model. In model V, the outer planet is introduced $\Delta_2 = 6\times10^5\,$yr after the inner one. This reduces the relative strength of the outer substructure by $ \sim 80\%$ compared to model IV. In model VI, the outer planet mass is tapered over $\tau_2 = 6\times10^5\,$yr. This reduces the outer rings flux by $\sim 40\%$ relative to the base model IV. Both of these models are within $3\sigma$ around the ALMA detection. However, the outer ring is not more symmetrical in either one of them.
        
        Finally, we present the result for a higher disk viscosity of $\alpha=10^{-3}$ in Fig. \ref{im_a3} (model VII). Motivated by the previous results, model VII used the self-similar solution initial profile and the fiducial mass tapering setup ($\Delta_{1,2}=0,\,\tau_{1,2}=16$ Kyr). This model produces an azimuthally symmetric ring. This can be explained as follows: While vortices can live long at the edge of a gap opened by a massive planet due to the Rossby wave instability in a disk with a low viscosity of $\alpha=10^{-4}$, they dissipate earlier with the higher viscosity $\alpha=10^{-3}$ (\citet{Lin_2011}). However, the minimum planetary mass needs to be increased in order to open a comparable gap in this higher-viscosity scenario. We find the mass estimate for the outer planet to be $3 M_{\rm{Jup}}$ (see Eq.~\eqref{mass_master_model_eq}). The mass of the inner planet remains unchanged ($8 M_{\rm{Jup}}$).
        
        The outer substructure in the synthetic observation of model VII agrees well with the ALMA observation because it does not display the asymmetry seen in previous models. Its normalized radial profile resides within a frame of $3\sigma$ around the detection. Consequently, model VII leads to the best agreement between our synthetic observation and the real one.
        
        To mimic how ALMA would see the disk that model VII generates, we took corresponding synthetic observation given by \texttt{RADMC-3D} and used the package \texttt{galario} to calculate the visibilities with the same uv-coverage as the observations. Fig.~\ref{im_a3}b shows the synthetic image obtained through radiative transfer and convolved with a Gaussian beam in the image space, as we did for all the other synthetic observation, and Fig.~\ref{im_a3}c shows the simulated observation using the visibilities, imaged with the same conditions as the observations. Both images are consistent with each another and reproduce the observed radial profile, proving the observability of the structures generated in our simulations.

\section{Discussion} \label{discussions}
    
    From the results above, it is clear that two planets are needed to explain the newly resolved outer ring in addition to the inner feature. None of the one-planet models of the disk (I and II) reproduce any strong visible substructures that would resemble the appearance of the outer ring, while the two-planet models feature two rings (e.g., VII). This conclusion is confirmed quantitatively by analyzing the errors of the normalized radial profiles: the models with a single planet have an above-significant deviation ($>3\sigma$) from the observation, while the models with two planets IV-VII have a deviation of less than $3\sigma$ in the radial profile. This finding is consistent with the results of \citet{p15,fedele2021}, who both concluded that the outer substructure is likely an indicator of a second companion.
    
    Additionally, we find that the initial density profile needs to have a stronger radial fall-off than the commonly assumed power law (Eq.~\ref{pow_eq}) in order to model the flux ratio of the inner and outer ring correctly. We have achieved this by using the self-similar solution from \citet{lynden-bell_pringle_1974} as initial condition.
    
    Our study also shows that changing the formation timing of the outer planet does affect the resulting radial profile and synthetic image (models IV, V, and VI), but all three models still reside within a $3\sigma$ range around the observation in the normalized radial profile. This means that the outer planets mass of $1.7 M_{\rm{Jup}}$ in model V, where the planet started to form very recently ($140\,$Kyr before the end of the simulation), is sufficient to successfully induce the outer substructure. This is much lower than previously estimated: \citet{p15} estimated that if the outer planet is much younger than the inner planet (1\,Myr younger), it should be more massive ($\gtrsim15\,M_{\rm{Jup}}$). This difference probably arises from the two different methods. \citet{p15} assumed an analytical solution to introduce the gaps in the gas surface density profile and ran dust evolution models considering that the gas density remains constant over time. In our current approach, however, we modeled gas and dust evolution simultaneously, but we neglected dust growth.
    
    Our study favors a \citet{s73} disk viscosity of $\alpha\gtrsim10^{-3}$  because we searched for an $\alpha$ that avoids asymmetric accumulations in the outer ring, which is successfully achieved in model VII (Fig.~\ref{im_a3}). Models of a disk with a lower viscosity ($\alpha=10^{-4}$) induce an asymmetric outer feature with a strong density accumulation that is absent in the ALMA image, and they also feature an accumulation of material around the position of the outer planet (see Fig.~\ref{im_stf}c). However, from the study of \citet{b20}, we suspect that this accumulation is an artifact of not simulating the accretion of the surrounding material onto the planet. This poses the question whether the shape of the outer ring itself would be as strongly affected by including accretion in the simulation as the gap. If this were the case, a simulation with a low viscosity $\alpha=10^{-4}$ might be able to produce a symmetrical feature with the outer planet mass estimate as low as $1.7\,M_{\rm{Jup}}$. Further simulations including the planet accretion process are thus needed to address this question. This lies beyond the scope of this paper, however.

    In a recent study, \citet{fedele2021} examined the HD\,100546 disk planet setup with smooth particle hydrodynamics (SPH) simulations. They qualitatively reproduced the disk appearance with an inner planet of final mass of $3.1\,M_{\rm{Jup}}$ at $15\,$au and an outer planet that reached $8.5\,M_{\rm{Jup}}$ at $110\,$au. While the number of planets and the positions are consistent with what we derived from the radial profile in Sect. \ref{planet_setups} (inner planet: $13\,$au; outer planet: $143\,$au), the mass estimates differ from our results (inner planet: $8\,M_{\rm{Jup}}$; outer planet: $3\,M_{\rm{Jup}}$ for $\alpha=10^{-3}$, or $1.7\,M_{\rm{Jup}}$ for $\alpha=10^{-4}$). The origin of these differences is unclear. One possible explanation is that for the outer planet mass, \citet{fedele2021} reported an estimate based on the the gap width with the same model as we used (Eq. \ref{mass_K_range_formula}), but fixed the proportionality constant to an average value while we chose the minimum possible mass; and they have a higher estimate of $100\,$au for the dust gap width than we do ($79\,$au). As a consequence, their estimated planet mass ($5.5\,M_{\rm{Jup}}$) is a higher than our estimate ($3\,M_{\rm{Jup}}$). 

     The results we deduced from submillimeter ALMA observations compare well to findings from observations at other wavelengths. \citet{Brittain_2014} hypothesized the presence of a planet in the HD\,100546 system with an orbit of ~15 AU from molecular emission line observations. This companion appears consistent with the inner planet of our best model (13 AU). Furthermore, \citet{Follette_2017} analyzed scattered-light observations of HD\,100546, revealing multiple spiral arms. They proposed that a planet of a few Jupiter masses at ~100 AU could explain these observed spiral structures. This planet roughly matches the outer planet from our model (at 143 AU).

\section{Conclusions} \label{conclusions}

    We aimed to constrain some of the properties of potential planets embedded in the HD100546\,circumstellar disk from the observed substructures. In particular, we were interested in constraining their masses and orbits. To do this, we analyzed recent ALMA observations of the system, and then matched them as closely as possible with our numerical model, iteratively improving the fit by systematically varying disk and planet parameters. In our numerical model, we first simulated the gas and dust in a protoplanetary disk with the multifluid code \texttt{FARGO3D}, and then rendered a synthetic observation from the results with \texttt{RADMC-3D}.
    
    Thereby, we showed that the recent ALMA observations of HD\,100546 can be reproduced by a protoplanetary disk model hosting two companions orbiting at separations of $13\,$au and $143\,$au. The best agreement between synthetic and real observations is reached with model VII, which uses the self-similar solution as initial condition and value of $\alpha = 10^{-3}$. This model matches the observation within $3\sigma$ of certainty in the normalized radial profile, and the resulting outer ring does not show azimuthal asymmetries. From the observed intensity profile, we estimate a mass of $8\,M_{\rm{Jup}}$ for the inner planet. This agrees with the results of other studies (\citet{p15}: $  \sim 10\,M_{\rm{Jup}}$), but it is higher than suggested by \citet{fedele2021} ($3.1\,M_{\rm{Jup}}$).
    
    In contrast to previous results, we conclude that the outer planets mass can be as low as $3\,M_{\rm{Jup}}$ for a disk viscosity of about $\alpha=10^{-3}$, and $1.7\,M_{\rm{Jup}}$ for a lower viscosity ($\alpha=10^{-4}$). Our results show that this companion can successfully reproduce the outer substructure, both in a model were it formed early in the disks lifetime and in a model where it formed very recently (as young as $140$ Kyr), equally well. This estimate for the outer planet mass is low in comparison to the previous studies of \citet{p15} ($\gtrsim15\,M_{\rm{Jup}}$ for a young planet, and $\lesssim 5\,M_{\rm{Jup}}$ for an older planet) and \citet{fedele2021} ($ 8.5\,M_{\rm{Jup}}$). We further find that to properly reproduce the brightness ratio of the inner and outer ring, the disks initial gas surface density profile needs to be modeled as a self-similar solution profile.
    
    Finally, the planet mass estimates need to be evaluated in terms of whether companions this massive might be directly visually detected by current instrumentation capabilities. \citet{p20} have deduced a lower limit for this in ALMA observation, depending on the star age. They found that the planets of this system would have to be more massive than $33\,M_{\rm{Jup}}$ for an age of $4\,$Myr to be detected by ALMA. This means that the detection of the proposed planets with $8\,M_{\rm{Jup}}$ and $3\,M_{\rm{Jup}}$ is still beyond our current ALMA capabilities. In addition, they are very difficult to directly image with instruments like SPHERE at the VLT  (\citet{boccaletti2020sphere}). Future high-resolution observations are therefore needed to prove or rule out the existence of embedded planets in the HD\,100546 protoplanetary disk.

\begin{acknowledgements}
    We would like to thank Camille Bergez-Casalou, Cornelis Dullemond and Marcelo Barraza for helpful discussions and technical insights. The Computational resources of the Max Planck Computing and Data Facility made this project possible and are gratefully acknowledged. We are also thankful for the support provided by the Alexander von Humboldt Foundation in the framework of the Sofja Kovalevskaja Award endowed by the Federal Ministry of Education and Research. We further acknowledge the efforts of Adriana Pohl, Akimasa Kataoka, and Cornelis Dullemond to motivate and propose the observation of the HD\,100546 system with ALMA. We thank Ruobing Dong for revising this manuscript and providing constructive feedback and suggestions.
    This paper makes use of the following ALMA data: ADS/JAO.ALMA\#2016.1.00497.S., and ADS/JAO.ALMA\#2015.1.00806.S. ALMA is a partnership of ESO (representing its member states), NSF (USA) and NINS (Japan), together with NRC (Canada), MOST and ASIAA (Taiwan), and KASI (Republic of Korea), in cooperation with the Republic of Chile. The Joint ALMA Observatory is operated by ESO, AUI/NRAO and NAOJ.
\end{acknowledgements}

\bibliographystyle{aa}
\bibliography{citations}

\begin{thebibliography}{59}
\expandafter\ifx\csname natexlab\endcsname\relax\def\natexlab#1{#1}\fi

\bibitem[{{ALMA Partnership} {et~al.}(2015){ALMA Partnership}, Brogan, Pérez,
  Hunter, Dent, Hales, Hills, Corder, Fomalont, Vlahakis, \&
  et~al.}]{alma_2015}
{ALMA Partnership}, Brogan, C.~L., Pérez, L.~M., {et~al.} 2015, The
  Astrophysical Journal, 808, L3

\bibitem[{{Andrews}(2020)}]{andrews2020}
{Andrews}, S.~M. 2020, \araa, 58, 483

\bibitem[{{Andrews} {et~al.}(2021){Andrews}, {Elder}, {Zhang}, {Huang},
  {Benisty}, {Kurtovic}, {Wilner}, {Zhu}, {Carpenter}, {P{\'e}rez}, {Teague},
  {Isella}, \& {Ricci}}]{andrews2021}
{Andrews}, S.~M., {Elder}, W., {Zhang}, S., {et~al.} 2021, \apj, 916, 51

\bibitem[{{Andrews} {et~al.}(2018){Andrews}, {Huang}, {P{\'e}rez}, {Isella},
  {Dullemond}, {Kurtovic}, {Guzm{\'a}n}, {Carpenter}, {Wilner}, {Zhang}, {Zhu},
  {Birnstiel}, {Bai}, {Benisty}, {Hughes}, {{\"O}berg}, \&
  {Ricci}}]{andrews2018}
{Andrews}, S.~M., {Huang}, J., {P{\'e}rez}, L.~M., {et~al.} 2018, \apjl, 869,
  L41

\bibitem[{Andrews {et~al.}(2018)Andrews, Huang, Pérez, Isella, Dullemond,
  Kurtovic, Guzmán, Carpenter, Wilner, Zhang, \& et~al.}]{Andrews_2018}
Andrews, S.~M., Huang, J., Pérez, L.~M., {et~al.} 2018, The Astrophysical
  Journal, 869, L41

\bibitem[{{Andrews} {et~al.}(2013){Andrews}, {Rosenfeld}, {Kraus}, \&
  {Wilner}}]{andrews2013}
{Andrews}, S.~M., {Rosenfeld}, K.~A., {Kraus}, A.~L., \& {Wilner}, D.~J. 2013,
  \apj, 771, 129

\bibitem[{{Ansdell} {et~al.}(2016){Ansdell}, {Williams}, {van der Marel},
  {Carpenter}, {Guidi}, {Hogerheijde}, {Mathews}, {Manara}, {Miotello},
  {Natta}, {Oliveira}, {Tazzari}, {Testi}, {van Dishoeck}, \& {van
  Terwisga}}]{ansdell2016}
{Ansdell}, M., {Williams}, J.~P., {van der Marel}, N., {et~al.} 2016, \apj,
  828, 46

\bibitem[{Ardila {et~al.}(2007)Ardila, Golimowski, Krist, Clampin, Ford, \&
  Illingworth}]{a7}
Ardila, D.~R., Golimowski, D.~A., Krist, J.~E., {et~al.} 2007, The
  Astrophysical Journal, 665, 512

\bibitem[{Bae {et~al.}(2019)Bae, Zhu, Baruteau, Benisty, Dullemond, Facchini,
  Isella, Keppler, Pérez, \& Teague}]{Bae_2019}
Bae, J., Zhu, Z., Baruteau, C., {et~al.} 2019, The Astrophysical Journal, 884,
  L41

\bibitem[{{Ben{\'\i}tez-Llambay} {et~al.}(2019){Ben{\'\i}tez-Llambay}, {Krapp},
  \& {Pessah}}]{2019fargomulti}
{Ben{\'\i}tez-Llambay}, P., {Krapp}, L., \& {Pessah}, M.~E. 2019, \apjs, 241,
  25

\bibitem[{Benitez-Llambay \& Masset(2016)}]{b16}
Benitez-Llambay, P. \& Masset, F.~S. 2016, The Astrophysical Journal Supplement
  Series, 223, 11

\bibitem[{Bergez-Casalou {et~al.}(2020)Bergez-Casalou, Bitsch, Pierens, Crida,
  \& Raymond}]{b20}
Bergez-Casalou, C., Bitsch, B., Pierens, A., Crida, A., \& Raymond, S.~N. 2020,
  Astronomy \& Astrophysics, 643, A133

\bibitem[{Boccaletti {et~al.}(2020)Boccaletti, Chauvin, Mouillet, Absil,
  Allard, Antoniucci, Augereau, Barge, Baruffolo, Baudino, Baudoz, Beaulieu,
  Benisty, Beuzit, Bianco, Biller, Bonavita, Bonnefoy, Bos, Bouret, Brandner,
  Buchschache, Carry, Cantalloube, Cascone, Carlotti, Charnay, Chiavassa,
  Choquet, Clenet, Crida, Boer, Caprio, Desidera, Desert, Delisle, Delorme,
  Dohlen, Doelman, Dominik, Orazi, Dougados, Doute, Fedele, Feldt, Ferreira,
  Fontanive, Fusco, Galicher, Garufi, Gendron, Ghedina, Ginski, Gonzalez,
  Gratadour, Gratton, Guillot, Haffert, Hagelberg, Henning, Huby, Janson, Kamp,
  Keller, Kenworthy, Kervella, Kral, Kuhn, Lagadec, Laibe, Langlois, Lagrange,
  Launhardt, Leboulleux, Coroller, Causi, Loupias, Maire, Marleau, Martinache,
  Martinez, Mary, Mattioli, Mazoyer, Meheut, Menard, Mesa, Meunier, Miguel,
  Milli, Min, Molliere, Mordasini, Moretto, Mugnier, Arena, Nardetto, Diaye,
  Nesvadba, Pedichini, Pinilla, Por, Potier, Quanz, Rameau, Roelfsema, Rouan,
  Rigliaco, Salasnich, Samland, Sauvage, Schmid, Segransan, Snellen, Snik,
  Soulez, Stadler, Stam, Tallon, Thebault, Thiebaut, Tschudi, Udry, van
  Holstein, Vernazza, Vidal, Vigan, Waters, Wildi, Willson, Zanutta, Zavagno,
  \& Zurlo}]{boccaletti2020sphere}
Boccaletti, A., Chauvin, G., Mouillet, D., {et~al.} 2020, SPHERE+: Imaging
  young Jupiters down to the snowline

\bibitem[{Brittain {et~al.}(2014)Brittain, Carr, Najita, Quanz, \&
  Meyer}]{Brittain_2014}
Brittain, S.~D., Carr, J.~S., Najita, J.~R., Quanz, S.~P., \& Meyer, M.~R.
  2014, The Astrophysical Journal, 791, 136

\bibitem[{Brown {et~al.}(2021)Brown, Vallenari, Prusti, de~Bruijne, Babusiaux,
  Biermann, Creevey, Evans, Eyer, \& et~al.}]{gaia_edr3_2021}
Brown, A. G.~A., Vallenari, A., Prusti, T., {et~al.} 2021, Astronomy \&
  Astrophysics, 650, C3

\bibitem[{{Casassus} \& {P{\'e}rez}(2019)}]{casassus_2019}
{Casassus}, S. \& {P{\'e}rez}, S. 2019, \apjl, 883, L41

\bibitem[{{Cieza} {et~al.}(2021){Cieza}, {Gonz{\'a}lez-Ruilova}, {Hales},
  {Pinilla}, {Ru{\'\i}z-Rodr{\'\i}guez}, {Zurlo}, {Casassus}, {P{\'e}rez},
  {C{\'a}novas}, {Arce-Tord}, {Flock}, {Kurtovic}, {Marino}, {Nogueira},
  {Perez}, {Price}, {Principe}, \& {Williams}}]{cieza2021}
{Cieza}, L.~A., {Gonz{\'a}lez-Ruilova}, C., {Hales}, A.~S., {et~al.} 2021,
  \mnras, 501, 2934

\bibitem[{Crida {et~al.}(2006)Crida, Morbidelli, \& Masset}]{c6}
Crida, A., Morbidelli, A., \& Masset, F. 2006, Icarus, 181, 587–604

\bibitem[{{Currie} {et~al.}(2015){Currie}, {Cloutier}, {Brittain}, {Grady},
  {Burrows}, {Muto}, {Kenyon}, \& {Kuchner}}]{currie2015}
{Currie}, T., {Cloutier}, R., {Brittain}, S., {et~al.} 2015, \apjl, 814, L27

\bibitem[{{Currie} {et~al.}(2014){Currie}, {Muto}, {Kudo}, {Honda}, {Brandt},
  {Grady}, {Fukagawa}, {Burrows}, {Janson}, {Kuzuhara}, {McElwain}, {Follette},
  {Hashimoto}, {Henning}, {Kandori}, {Kusakabe}, {Kwon}, {Mede}, {Morino},
  {Nishikawa}, {Pyo}, {Serabyn}, {Suenaga}, {Takahashi}, {Wisniewski}, \&
  {Tamura}}]{currie2014}
{Currie}, T., {Muto}, T., {Kudo}, T., {et~al.} 2014, \apjl, 796, L30

\bibitem[{Dodson-Robinson \& Salyk(2011)}]{Dodson_Robinson_2011}
Dodson-Robinson, S.~E. \& Salyk, C. 2011, The Astrophysical Journal, 738, 131

\bibitem[{Dong {et~al.}(2017)Dong, Li, Chiang, \& Li}]{Dong_2017}
Dong, R., Li, S., Chiang, E., \& Li, H. 2017, The Astrophysical Journal, 843,
  127

\bibitem[{Dong {et~al.}(2018)Dong, Li, Chiang, \& Li}]{Dong_2018}
Dong, R., Li, S., Chiang, E., \& Li, H. 2018, The Astrophysical Journal, 866,
  110

\bibitem[{{Dorschner} {et~al.}(1995){Dorschner}, {Begemann}, {Henning},
  {Jaeger}, \& {Mutschke}}]{d95}
{Dorschner}, J., {Begemann}, B., {Henning}, T., {Jaeger}, C., \& {Mutschke}, H.
  1995, \aap, 300, 503

\bibitem[{Dullemond \& Dominik(2004)}]{d4}
Dullemond, C.~P. \& Dominik, C. 2004, Astronomy \& Astrophysics, 421,
  1075–1086

\bibitem[{{Dullemond} {et~al.}(2012){Dullemond}, {Juhasz}, {Pohl}, {Sereshti},
  {Shetty}, {Peters}, {Commercon}, \& {Flock}}]{d12}
{Dullemond}, C.~P., {Juhasz}, A., {Pohl}, A., {et~al.} 2012, {RADMC-3D: A
  multi-purpose radiative transfer tool}

\bibitem[{{Facchini} {et~al.}(2018){Facchini}, {Pinilla}, {van Dishoeck}, \&
  {de Juan Ovelar}}]{2018facchini}
{Facchini}, S., {Pinilla}, P., {van Dishoeck}, E.~F., \& {de Juan Ovelar}, M.
  2018, \aap, 612, A104

\bibitem[{Fedele {et~al.}(2021)Fedele, Toci, Maud, \& Lodato}]{fedele2021}
Fedele, D., Toci, C., Maud, L., \& Lodato, G. 2021, ALMA 870 $\mu$m continuum
  observations of HD 100546. Evidence of a giant planet on a wide orbit

\bibitem[{Follette {et~al.}(2017)Follette, Rameau, Dong, Pueyo, Close,
  Duchêne, Fung, Leonard, Macintosh, Males, \& et~al.}]{Follette_2017}
Follette, K.~B., Rameau, J., Dong, R., {et~al.} 2017, The Astronomical Journal,
  153, 264

\bibitem[{Fung \& Chiang(2016)}]{Fung_2016}
Fung, J. \& Chiang, E. 2016, The Astrophysical Journal, 832, 105

\bibitem[{Hildebrand(1983)}]{h83}
Hildebrand. 1983, Quarterly Journal of the Royal Astronomical Society, 24, 267

\bibitem[{{Jorsater} \& {van Moorsel}(1995)}]{jorsater1995}
{Jorsater}, S. \& {van Moorsel}, G.~A. 1995, \aj, 110, 2037

\bibitem[{{Lin} \& {Papaloizou}(2011)}]{Lin_2011}
{Lin}, M.-K. \& {Papaloizou}, J. C.~B. 2011, \mnras, 415, 1445

\bibitem[{Lodato {et~al.}(2019)Lodato, Dipierro, Ragusa, Long, Herczeg,
  Pascucci, Pinilla, Manara, Tazzari, Liu, \& et~al.}]{l19}
Lodato, G., Dipierro, G., Ragusa, E., {et~al.} 2019, Monthly Notices of the
  Royal Astronomical Society, 486, 453–461

\bibitem[{{Long} {et~al.}(2018){Long}, {Pinilla}, {Herczeg}, {Harsono},
  {Dipierro}, {Pascucci}, {Hendler}, {Tazzari}, {Ragusa}, {Salyk}, {Edwards},
  {Lodato}, {van de Plas}, {Johnstone}, {Liu}, {Boehler}, {Cabrit}, {Manara},
  {Menard}, {Mulders}, {Nisini}, {Fischer}, {Rigliaco}, {Banzatti}, {Avenhaus},
  \& {Gully-Santiago}}]{long2018}
{Long}, F., {Pinilla}, P., {Herczeg}, G.~J., {et~al.} 2018, \apj, 869, 17

\bibitem[{{Lynden-Bell} \& {Pringle}(1974)}]{lynden-bell_pringle_1974}
{Lynden-Bell}, D. \& {Pringle}, J.~E. 1974, \mnras, 168, 603

\bibitem[{Masset(2000)}]{m0}
Masset, F. 2000, Astronomy and Astrophysics Supplement Series, 141, 165–173

\bibitem[{{Mathis} {et~al.}(1977){Mathis}, {Rumpl}, \&
  {Nordsieck}}]{mathis_1977}
{Mathis}, J.~S., {Rumpl}, W., \& {Nordsieck}, K.~H. 1977, \apj, 217, 425

\bibitem[{{Montesinos} {et~al.}(2015){Montesinos}, {Cuadra}, {Perez},
  {Baruteau}, \& {Casassus}}]{montesinos2015}
{Montesinos}, M., {Cuadra}, J., {Perez}, S., {Baruteau}, C., \& {Casassus}, S.
  2015, \apj, 806, 253

\bibitem[{{P{\'e}rez} {et~al.}(2020){P{\'e}rez}, {Casassus}, {Hales}, {Marino},
  {Cheetham}, {Zurlo}, {Cieza}, {Dong}, {Alarc{\'o}n}, {Ben{\'\i}tez-Llambay},
  {Fomalont}, \& {Avenhaus}}]{perez_2020}
{P{\'e}rez}, S., {Casassus}, S., {Hales}, A., {et~al.} 2020, \apjl, 889, L24

\bibitem[{P{\'{e}}rez {et~al.}(2020)P{\'{e}}rez, Casassus, Hales, Marino,
  Cheetham, Zurlo, Cieza, Dong, Alarc{\'{o}}n, Ben{\'{\i}}tez-Llambay,
  Fomalont, \& Avenhaus}]{p20}
P{\'{e}}rez, S., Casassus, S., Hales, A., {et~al.} 2020, The Astrophysical
  Journal, 889, L24

\bibitem[{{Pinilla} {et~al.}(2012){Pinilla}, {Benisty}, \& {Birnstiel}}]{pi1}
{Pinilla}, P., {Benisty}, M., \& {Birnstiel}, T. 2012, \aap, 545, A81

\bibitem[{Pinilla {et~al.}(2012)Pinilla, Benisty, \& Birnstiel}]{Pinilla_2012}
Pinilla, P., Benisty, M., \& Birnstiel, T. 2012, Astronomy \& Astrophysics,
  545, A81

\bibitem[{Pinilla {et~al.}(2015)Pinilla, Birnstiel, \& Walsh}]{p15}
Pinilla, P., Birnstiel, T., \& Walsh, C. 2015, Astronomy \& Astrophysics, 580,
  A105

\bibitem[{{Pinilla} {et~al.}(2015){Pinilla}, {de Juan Ovelar}, {Ataiee},
  {Benisty}, {Birnstiel}, {van Dishoeck}, \& {Min}}]{pi2}
{Pinilla}, P., {de Juan Ovelar}, M., {Ataiee}, S., {et~al.} 2015, \aap, 573, A9

\bibitem[{Pinilla {et~al.}(2016)Pinilla, Flock, Ovelar, \&
  Birnstiel}]{Pinilla_2016}
Pinilla, P., Flock, M., Ovelar, M. d.~J., \& Birnstiel, T. 2016, Astronomy \&
  Astrophysics, 596, A81

\bibitem[{{Pinilla} {et~al.}(2018){Pinilla}, {Natta}, {Manara}, {Ricci},
  {Scholz}, \& {Testi}}]{pinilla2018}
{Pinilla}, P., {Natta}, A., {Manara}, C.~F., {et~al.} 2018, \aap, 615, A95

\bibitem[{Quanz {et~al.}(2015)Quanz, Amara, Meyer, Girard, Kenworthy, \&
  Kasper}]{q15}
Quanz, S.~P., Amara, A., Meyer, M.~R., {et~al.} 2015, The Astrophysical
  Journal, 807, 64

\bibitem[{Quanz {et~al.}(2013)Quanz, Amara, Meyer, Kenworthy, Kasper, \&
  Girard}]{q13}
Quanz, S.~P., Amara, A., Meyer, M.~R., {et~al.} 2013, The Astrophysical
  Journal, 766, L1

\bibitem[{{Rice} {et~al.}(2006){Rice}, {Armitage}, {Wood}, \& {Lodato}}]{ri}
{Rice}, W.~K.~M., {Armitage}, P.~J., {Wood}, K., \& {Lodato}, G. 2006, \mnras,
  373, 1619

\bibitem[{Rosotti {et~al.}(2016)Rosotti, Juhasz, Booth, \&
  Clarke}]{rosotti2016}
Rosotti, G.~P., Juhasz, A., Booth, R.~A., \& Clarke, C.~J. 2016, Monthly
  Notices of the Royal Astronomical Society, 459, 2790

\bibitem[{{Shakura} \& {Sunyaev}(1973)}]{s73}
{Shakura}, N.~I. \& {Sunyaev}, R.~A. 1973, \aap, 500, 33

\bibitem[{{Stammler} {et~al.}(2017){Stammler}, {Birnstiel}, {Pani{\'c}},
  {Dullemond}, \& {Dominik}}]{st17}
{Stammler}, S.~M., {Birnstiel}, T., {Pani{\'c}}, O., {Dullemond}, C.~P., \&
  {Dominik}, C. 2017, \aap, 600, A140

\bibitem[{{Suriano} {et~al.}(2019){Suriano}, {Li}, {Krasnopolsky}, {Suzuki}, \&
  {Shang}}]{su193d}
{Suriano}, S.~S., {Li}, Z.-Y., {Krasnopolsky}, R., {Suzuki}, T.~K., \& {Shang},
  H. 2019, \mnras, 484, 107

\bibitem[{Walsh {et~al.}(2014)Walsh, Juh{\'{a}}sz, Pinilla, Harsono, Mathews,
  Dent, Hogerheijde, Birnstiel, Meeus, Nomura, Aikawa, Millar, \&
  Sandell}]{w14}
Walsh, C., Juh{\'{a}}sz, A., Pinilla, P., {et~al.} 2014, The Astrophysical
  Journal, 791, L6

\bibitem[{{Weber} {et~al.}(2019){Weber}, {P{\'e}rez}, {Ben{\'\i}tez-Llambay},
  {Gressel}, {Casassus}, \& {Krapp}}]{2019diffusionfargo}
{Weber}, P., {P{\'e}rez}, S., {Ben{\'\i}tez-Llambay}, P., {et~al.} 2019, \apj,
  884, 178

\bibitem[{Wichittanakom {et~al.}(2020)Wichittanakom, Oudmaijer, Fairlamb,
  Mendigutía, Vioque, \& Ababakr}]{Wichittanakom_2020}
Wichittanakom, C., Oudmaijer, R.~D., Fairlamb, J.~R., {et~al.} 2020, Monthly
  Notices of the Royal Astronomical Society, 493, 234–249

\bibitem[{Zhang {et~al.}(2016)Zhang, Bergin, Blake, Cleeves, Hogerheijde,
  Salinas, \& Schwarz}]{Zhang_2016}
Zhang, K., Bergin, E.~A., Blake, G.~A., {et~al.} 2016, The Astrophysical
  Journal, 818, L16

\bibitem[{{Zhu} \& {Stone}(2014)}]{zu3}
{Zhu}, Z. \& {Stone}, J.~M. 2014, \apj, 795, 53

\end{thebibliography}

\appendix
    \section{Observation details}
    
    In table \ref{tab:obs_log} we present the details of the Atacama Large Millimeter/sub-millimeter Array (ALMA) observations that were used in this work.
    
        \begin{table*}
            \begin{center}
            \caption{Summary of ALMA observations for HD\,100546.}
                \begin{tabular}{l c c c c c}
                    \hline
                    \hline
                    Program ID     &  Obs. Date  & Exp. Time & N$^\circ$ & Baselines & Configuration  \\
                                   &             & (min)     & Antennas  & (m)       &                \\
                    \hline
                    2015.1.00934.S &  2015-12-02 & 27.8      &  36       & 17.4 - 10803.7 & Extended (LB)  \\
                    2016.1.01511.S &  2016-10-26 & 69.4      &  43       & 15.1 - 1124.3  & Compact (SB)   \\
                                   &  2017-04-23 & 104.1     &  44       & 15.1 - 460.0   & Compact (SB)   \\
                    \hline
                \end{tabular}
            \end{center}
            
            \label{tab:obs_log}
        \end{table*}

    \section{Convergence and resolution tests} \label{conv_behav}
    
\begin{figure*}
        \centering
        \includegraphics[width=0.96\linewidth]{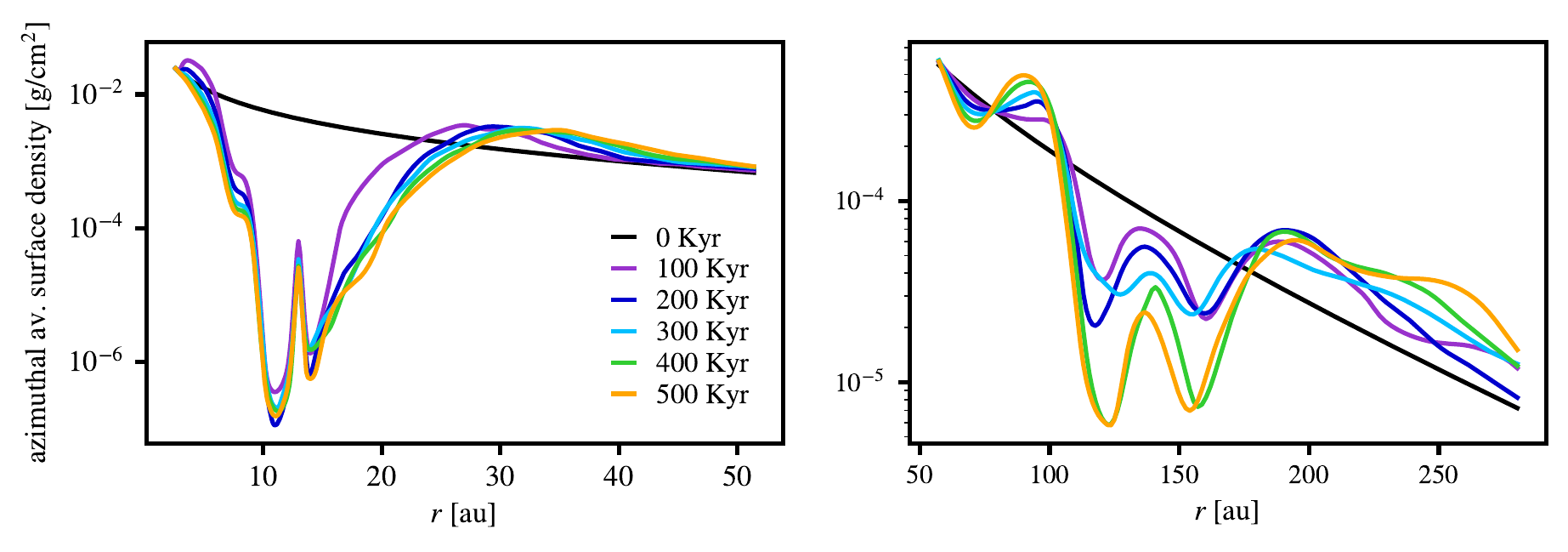}
        \caption{Evolution of the azimuthally averaged surface density profile, for the dust fluid with a grain size of $a_{\rm{dust}}=5\mu$m, from simulation IV.}
        \label{cp}
    \end{figure*}

    Our \texttt{FARGO3D} simulations were run for a shorter time than the age of the HD\,100546 system (simulation time: $ 0.74\,$Myr, estimated system age: $4.8\,\rm{Myr}$, \citet{Wichittanakom_2020}). This was done under the assumption that the disk model reaches a quasi-steady state within the simulated time. To confirm this assumption, we illustrate the temporal convergence behavior in the gap formation process of our model in Fig.~\ref{cp} and show that our gas surface density does not vary significantly from $200\,$Kyr.
    
\begin{figure*}
        \centering
        \includegraphics[width=0.96\linewidth]{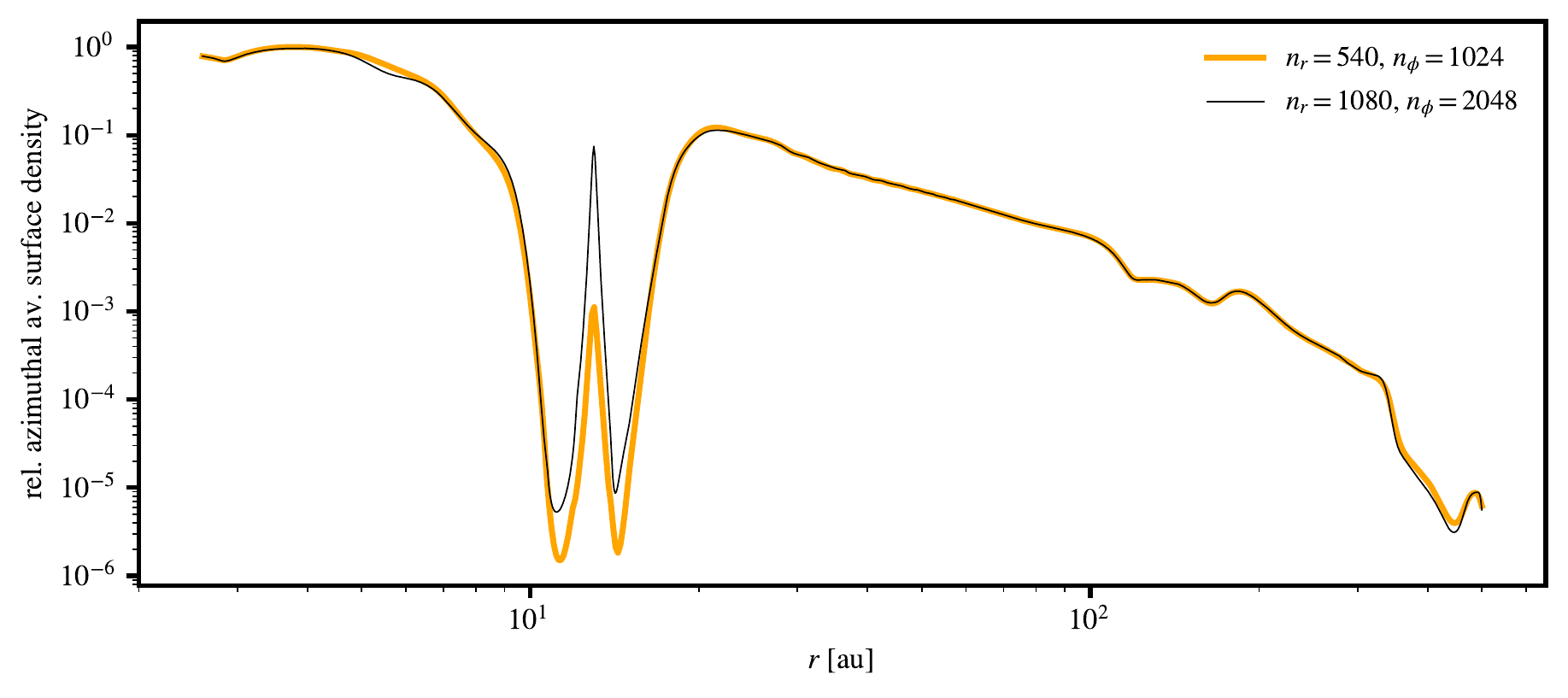}
        \caption{Radial profiles of the surface density results after 1000 inner planet orbits for the resolution used in all the simulations presented in the paper ($n_r=540$, $n_\phi=1024$) compared to a higher resolution ($n_r=1080$, $n_\phi=2048$). The results are for the dust fluid with a grain size of $a_{\rm{dust}}=5\mu$m with simulation setup VII in both cases.}
        \label{restest_gas}
    \end{figure*}

    Furthermore , we carried out all of our \texttt{FARGO3D} simulations on a two-dimensional grid that resolved one-tenth of the pressure scale height. The consequent dimensions of the grid were $n_r=540$ radial and $n_\phi=1024$ azimuthal cells. To show that this resolution is sufficient and does not affect the results, we present a comparison of model VII and a simulation with the identical setup, but resolving one-twentieth scale height ($n_r=1080$, $n_\phi=2048$) in Fig.~\ref{restest_gas}. The result does not change significantly by the higher resolution. The strongest deviation arises with the nonphysical material accumulation at the location of the inner planet (13\,au), where the peak is higher.

\end{document}